**Mechanism of Spin-Orbit Torques in Platinum Oxide Systems**

*Jayshankar Nath*, Alexandru Vladimir Trifu, Mihai Sebastian Gabor, Ali Hallal, Stephane Auffret, Sebastien Labau, Aymen Mahjoub, Edmond Chan, Avinash Kumar Chaurasiya, Amrit Kumar Mondal, Haozhe Yang, Eva Schmoranzerova, Mohamed Ali Nsibi, Isabelle Joumard, Anjan Barman, Bernard Pelissier, Mairbek Chshiev, Gilles Gaudin, and Ioan Mihai Miron**

Dr. J. Nath, Dr. A. V. Trifu, Dr. A. Hallal, S. Auffret, E. Chan, Dr. H. Yang, Dr. E. Schmoranzerova, Dr. M. A. Nsibi, Dr. I. Joumard, Prof. M. Chshiev, Dr. G. Gaudin, Dr. I. M. Miron
SPINTEC, UMR-8191,
CNRS/CEA/Universite Grenoble Alpes,
38054 Grenoble, France
E-mail: jsnath@ucla.edu, mihai.miron@cea.fr

Dr. M. S. Gabor
C4S, Physics and Chemistry Department,
Technical University of Cluj-Napoca,
400114 Cluj-Napoca, Romania

Dr. S. Labau, Dr. A. Mahjoub, Dr. B. Pelissier
Universite Grenoble Alpes,
CNRS, CEA/LETI Minatec, LTM,
38054 Grenoble, France

A. K. Chaurasiya, A. K. Mondal, Prof. A. Barman
Department of Condensed Matter Physics and Material Sciences,
S. N. Bose National Centre for Basic Sciences
700106 Kolkata, India

Prof. M. Chshiev
Institut Universitaire de France (IUF)
75231 Paris, France

**Spin-Orbit Torque (SOT) Magnetic Random-Access Memories (MRAM) have shown promising results towards the realization of fast, non-volatile memory systems. Oxidation of the heavy-metal (HM) layer of the SOT-MRAM has been proposed as a method to increase its energy efficiency. But the results are widely divergent due to the difficulty in controlling the HM oxidation because of its low enthalpy of formation. Here, we reconcile these differences by performing a gradual oxidation procedure, which allows correlating the chemical structure to the physical properties of the stack. As an HM layer, we chose Pt because of the strong SOT and the low enthalpy of formation of its oxides. We find evidence of an oxide inversion layer at the FM/HM interface: the oxygen is drawn into the FM, while the HM remains metallic near the interface. We further demonstrate that the oxygen migrates in the volume of the FM layer rather than being concentrated at the interface. Consequently, we find that the intrinsic magnitude of the SOT is unchanged compared to the fully metallic structure. The previously reported apparent increase of SOTs is not intrinsic to platinum oxide and instead arises from systemic changes produced by oxidation.**

## 1. Introduction

The process of transfering data between the memory and logic elements of a computer is one of the most energy intensive steps and is a bottleneck in terms of speed. Before moving to unconventional architectures such as “logic-in-memory” or “machine learning systems”, the most direct solution seems to be to include non-volatility in the memory hierarchy. For that, the SOT-MRAM [1] is the only non volatile memory technology that can function at the speed of the processors (at least at GHz). This memory is targeted towards replacing the current Complementary Metal-Oxide-Semiconductor (CMOS) based Static RAMs (SRAM) and Dynamic RAMs (DRAM), with a greater emphasis on key metrics such as energy efficiency, non-volatility, speed, size, and endurance.

In SOT-MRAM, a current injected into the HM layer with high Spin-Orbit Coupling (SOC) gives rise to the Damping-Like (DL) and the Field-Like (FL) torques. These torques act on the magnetization of the adjacent FM layer to switch it during the write operation. [2,3] The main figure of merit of SOT-MRAM is the switching current, which dictates the power consumption as well as the footprint of the transistors that drives the memory cell. [4]

Numerous works have been published, engineering either the bulk of the HM or the FM/HM interface to reduce the writing current: from using resistive HMs [3,5–7] to alloying [8,9] in the case of bulk methods and using insertion layers [10–12] and spin sinks [13,14] in case of interfacial methods. One particularly attractive method to enhance the SOT is to oxidize the HM. [15–19] Its main advantage is that the SOT created at the Oxide/FM interface could be stronger than the SOT from the HM, while also consuming less current, due to the insulating nature of the oxide. While most of the published works on this topic report an increase of the SOT efficacy, there is an ongoing debate as some other studies do not observe such a clear improvement. [20] This ambiguity in the experimental results published so far necessitates a clear study of HM oxide systems, as this issue is highly relevant from a technological perspective in the development of SOT-MRAMs.

## 2. Generation of SOTs by the oxidized platinum layer

In this work, we study the SOTs generated by oxidizing the platinum layer in a Ta(3 nm)/Cu(1 nm)/Co(2 nm)/Pt(4 - 1 nm) multilayer stack to determine the exact contribution of oxidation in these structures. The sample stack is shown in **Figure 1** (a). Here, the Cu/Co/Pt forms the inversion asymmetric tri-layer, while the Ta acts as a seed layer. The torques generated by Ta are measured and subtracted from the results. Unlike the bulk oxidation of the HM layer during deposition utilized in the previous works, [16–20] the wedge of platinum creates a continuous gradient of oxidation, while all the other properties of the samples remain unchanged. This specificity of the wedge ensures that all our samples are deposited at the same time, avoiding any material differences from sequential deposition runs. The sample was oxidized post-deposition [21] and was coated with Poly(methyl methacrylate) (PMMA) to avoid further atmospheric oxidation.

Devices were fabricated along the Pt wedge and the SOTs were measured using the 2$^{nd}$ harmonic torque measurement technique. [22] The DL fields thus extracted and normalized to the applied current are plotted in Figure 1 (b). At higher thicknesses of Pt, the unoxidized (UO) and the oxidized (OX) samples generate the same amount of torque. However, at lower thicknesses, the OX samples exhibit an increase in SOTs. This increase, also observed in previous studies, was interpreted as an indicator of SOTs arising from FM/HM interfacial oxidation. Moreover, since oxidation also increases the perpendicular magnetic anisotropy of the FM layer, novel SOT-MRAM devices can be imagined with the FM layer sandwiched between a top MgO layer and a bottom oxidized Pt layer, resulting in enhanced SOTs and perpendicular magnetic anisotropy.

For these reasons, it is critical to determine the precise role of oxygen at the FM/Pt interface. The first step is to determine the distribution of oxygen in the stack and its effect on the physical and electrical characteristics of the Pt layer. This can be ascertained by using the Angle-Resolved X-ray Photoelectron Spectroscopy (AR-XPS) technique plotted in Figure 1 (c) and (d). We selected two OX samples at 2 nm and 1.6 nm of Pt thickness, corresponding to either a negligible effect or a significant increase of SOTs, respectively. Figure 1 (c) displays the Pt spectra of the OX(2) and OX(1.6) samples. Both these samples indicate the presence of excited states of Pt, corresponding to its oxidized states, namely the $Pt^{2+}$ and $Pt^{4+}$. However, comparing the area under the peaks of $Pt^{0+}$ to $Pt^{2+}$ and $Pt^{4+}$, it is evident that OX(1.6) is more strongly oxidized owing to its thinner Pt thickness. Figure 1 (d) displays the Co spectra of these two samples. OX(2) doesn't show any excited states, indicating that the Co is unoxidized. However, OX(1.6) indicates the presence of the excited states corresponding to the oxides of Co, indicating that the oxygen in this sample has reached the Co/Pt interface. The correlation between the interfacial oxidation and the increase of SOTs could indicate an enhancement of the Rashba effect [23] as well as a change of the spin mixing conductance of this interface.

To test this hypothesis, we performed the Ferromagnetic Resonance (FMR) measurements on the samples, allowing us to determine the effective spin-mixing conductance, $g_{eff}$, plotted in Figure 1 (e). [21] This plot indicates that there is an enhancement of spin-mixing conductance at the interface, with interfacial oxidation. Hence, the increase of SOTs, which is an interfacial effect, could also arise from the enhancement of spin-mixing conductance at the interface. [13,24]

Up to this point, the results that we find are in agreement with previous studies, [17,18] reporting an enhancement of the SOT as well as a change of the spin mixing conductance. However, these conclusions are reached with the assumption that the Pt oxide remains stable while in contact with the metallic Co, which is not compatible with the relative enthalpy of formation of these oxides. It is energetically more favorable to form oxides of Co (CoO: -237.9 $kJmol^{-1}$; $Co_3O_4$: -891 $kJmol^{-1}$) [25] rather than oxides of Pt ($PtO_2$: -80 $kJmol^{-1}$; $Pt_3O_4$: -163 $kJmol^{-1}$). [26] Hence, we need to determine the exact distribution of the oxygen inside the functional stack to understand how the oxidation affects the magnetic properties as well as the electric current flow. We will then correct the measured values of the SOTs to include all the changes of the physical properties.

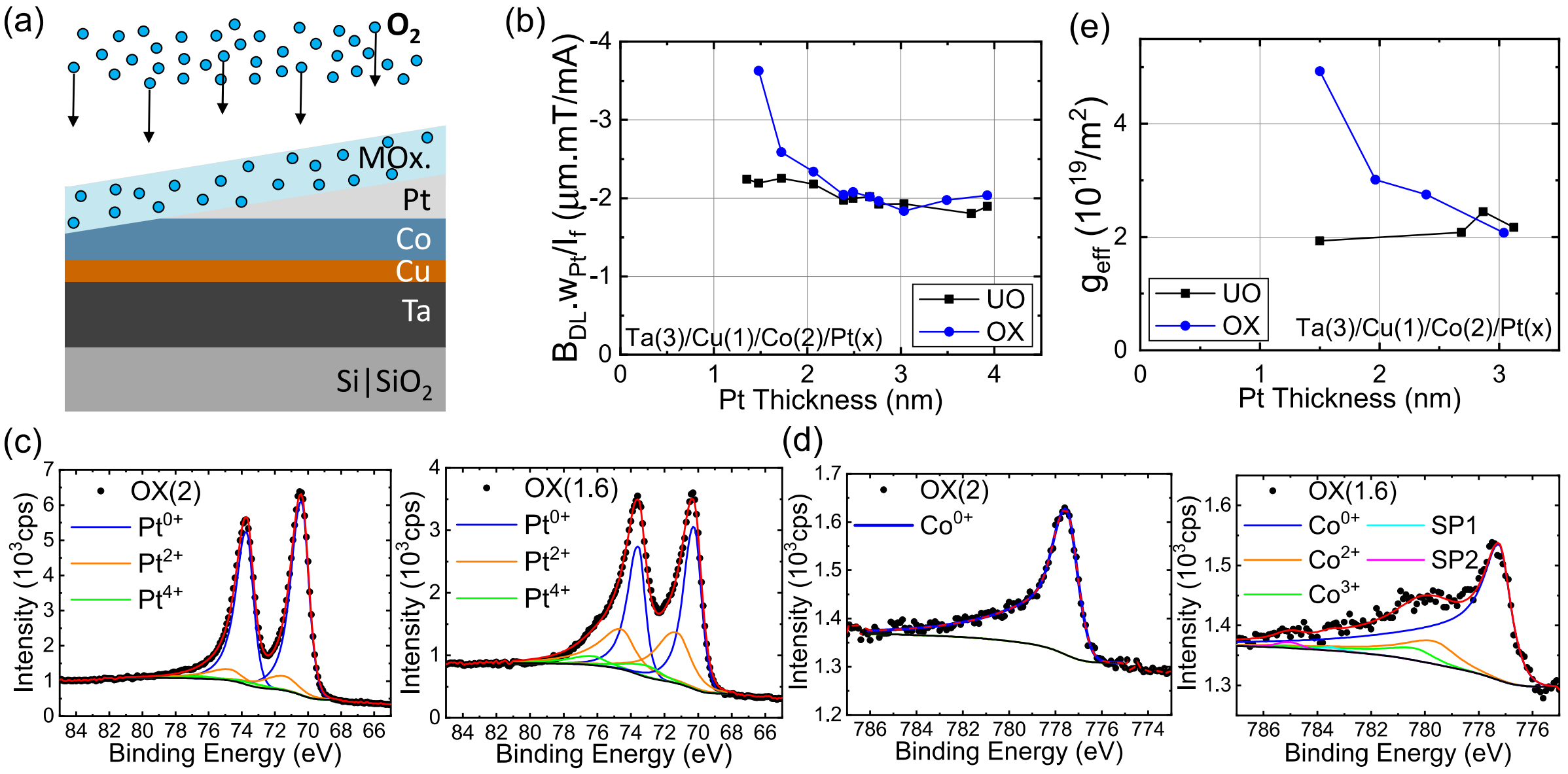

**Figure 1.** (a) Schematics of the multilayer stack of Ta(3 nm)/Cu(1 nm)/Co(2 nm)/Pt(4-1 nm). (b) DL field normalized to the applied current. (c) pAR-XPS Pt spectra (75° acquisition angle) of the OX(2 nm) (left) and OX(1.6 nm) (right) samples. (d) Co spectra (75° acquisition angle) of the OX(2 nm) (left) and OX(1.6 nm) (right). In (c) and (d), the black curve corresponds to the baseline spectra and the red line corresponds to the data fit. (e) Effective spin-mixing conductance as a function of Pt thickness. UO and OX refer to unoxidized and oxidized samples respectively. Samples UO (1.5 nm, 2.9 nm) of (e) were deposited separately.

## 3. Distribution of oxygen in the material stack

In 2nd harmonic torque measurements, we are comparing the action of current to that of the magnetic field. Hence, if the saturation magnetization, $M_S$, varies between the samples, then the extracted SOT field needs to be scaled accordingly. We measured the saturation magnetization using a Vibrating Sample Magnetometry (VSM). These data, after correcting for the Co thickness variation across the wafer, [21] are plotted in **Figure 2** (a). Above 1.8nm of Pt thickness, there is negligible difference between the $M_S$ of the samples. However, below this thickness of Pt, there is a significant drop in $M_S$, indicative of Co oxidation.

When a current is applied in a SOT device, it redistributes itself into the various metallic layers depending on their relative resistivities. However, the majority of the SOTs is generated in the functional Pt layer. To determine the current flowing within the Pt layer, we need to quantify the conductance of the layers. These were measured using the 4-point resistance measurement technique and the data fit with the Fuchs-Sondheimer (FS) model, [21,27,28] with results presented in Figure 2 (b). Because the Pt thicknesses here are smaller than the mean free path of electrons in the metal, we cannot extract meaningful values for all the physical parameters in the model. However, because the model fits the data sufficiently well, by extrapolating to zero thickness, we can determine the conductance of the layers underneath Pt. [21] The conductance of the OX samples, plotted in blue, follows the FS model up to around 1.8 nm thickness of Pt. However, below this thickness, it deviates from the model. This can occur due to oxygen migration into the Co layer, increasing the resistivity of this layer. Hence, there is a loss of conductance in a very conductive layer of the stack, resulting in the sharp drop at lower thicknesses of Pt. This is also consistent with the magnetization data, exhibiting a loss of magnetic volume around the same thickness of Pt. Moreover, the OX conductance curve crosses the solid red line, which is the limiting case of having no Pt layer. Hence, it is evident that the oxygen migrates from the platinum oxide into the Co layer. As this oxidation of the Co layer occurs before the complete oxidation of the Pt layer, it is indicative of an oxide inversion at the interface, wherein, at lower thicknesses of Pt, oxygen gets pumped into the Co layer, leaving the Pt metallic at the interface.

Based on these data, we propose a generic oxidation model shown in Figure 2 (c). The plasma oxidizes the top of the Pt layer, [29] enhancing its resistivity. This increase is equivalent to an effective insulating layer of 0.4 nm thick. [21] Moreover, if the Pt is sufficiently thin, the oxygen is not just "pushed" into the Co layer, the Co "attracts" the oxygen from Pt, leaving the interface metallic and hence conductive. This model is different from other works that either assumed a completely oxidized HM layer [17,18] or a completely conducting HM layer. [20] The effect of the metallic Pt at the interface needs to be taken into account while quantifying the SOTs.

To verify this model, we performed ab-initio calculations, [21] in order to determine the energy of the system as the oxygen is placed at different locations in the lattice, as shown in Figure 2 (d). Here, a plane of oxygen atoms is placed in the lattice along the X-Y plane, and the total energy of the system is calculated by relaxing the structure. By moving this plane of oxygen atoms along the Z direction, we can determine the energy cost of placing oxygen at different locations in Pt and Co. It is evident from this figure that it is energetically more favorable for the oxygen to remain in the Co layer than in the Pt layer. This is also consistent with the enthalpies of formation, mentioned earlier. Hence, near the interface, the oxygen atoms migrate from the Pt layer into the Co layer most likely via grain boundaries.

Now that the distribution of oxygen in Pt has been established, we need to determine the same for the Co layer. We consider the two extreme possibilities: Co oxidized only at the interface or Co oxidized in the bulk, within its entire thickness. In the first case, the reduction of the magnetic moment (Figure 2a) is due to the formation of a magnetic dead layer, while in the second case the thickness is maintained but the $M_S$ is reduced. To find out which of these scenarios is more realistic, we compare the interfacial anisotropy of our samples in both cases. The differences in $M_S$ would be reflected in the interfacial anisotropy, as

$$B_K^{int} = \mu_0 M_S + B_K \tag{1}$$

Here, the perpendicular interfacial anisotropy, $B_K^{int}$, is the difference between the $M_S$, which holds the magnetization of Co in the plane of the sample, and $B_K$, the anisotropy value obtained from the transport measurements of the Anomalous Hall Effect (AHE). These values are plotted in Figure 2 (e). [21] We find that, within the assumption of dead layer formation, the interfacial anisotropy remains much larger than in the case of uniform oxidation, indicating that the latter is the most plausible scenario. Indeed, it is well established by previous studies of optimized oxidation in Magnetic Tunnel Junctions (MTJ) [30–35] that as the capping layer (Al, [30–32] Mg, [32,33] or Ta [31,32]) is over-oxidized, and the oxygen starts penetrating the Co layer, the perpendicular anisotropy is strongly reduced. [34–36] This is also consistent with the fact that at these thicknesses, Co oxide does not form a passivating layer, leading to bulk oxidation. [37] Hence, in our oxidation model, once the oxygen is pumped into the Co layer, it diffuses into the bulk reducing the $M_S$ of the layer and does not form just an interfacial Co oxide layer.

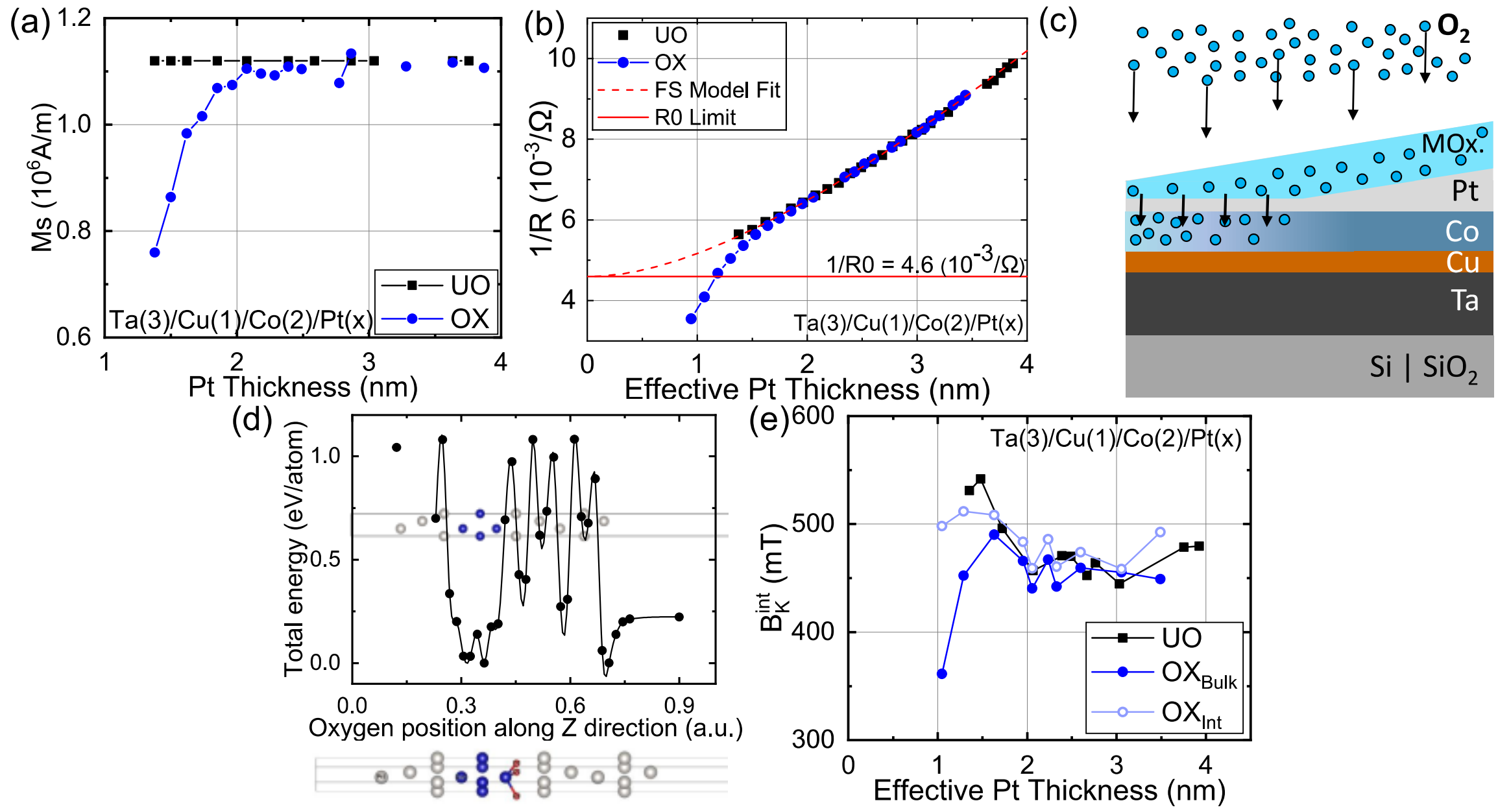

**Figure 2.** (a) Saturation magnetization of UO and OX samples. (b) Dependence of the conductance on the effective platinum thickness. The dashed line corresponds to the Fuchs-Sondheimer fit and the red line corresponds to the conductance of the layers underneath Pt. (c) Schematic of the proposed model illustrating the oxygen pumping into the Co layer. (d) The energy of the system (black dots) depending on the placement of the oxygen atoms in the lattice, determined using ab-initio calculations. The blue dots correspond to the Co atoms and the white dots correspond to the Pt atoms in the lattice. The bottom frame corresponds to the scenario when oxygen (red dots) is placed at the Co/Pt interface. (e) Perpendicular interfacial anisotropy determined from AHE measurements.

## 4. Quantifying SOTs based on the multi-layer oxidation model

Using the developed oxidation model, the torques are normalized to the current density in platinum and the electric field across it. These are plotted in terms of the corresponding DL-SOT efficiencies as a function of effective Pt thickness in **Figure 3** (a) and (b) respectively. No intrinsic increase in SOTs is observed after thorough corrections and normalizations. Comparing with published works, [17,18] there is indeed an apparent increase in SOTs at the system level. But after careful consideration of all the measurements, we conclude that there is no new property arising from Pt oxide. Instead, Co is pumping oxygen from the Pt layer, keeping the interface metallic. Moreover, regardless of the type of normalization of SOT efficiency, we obtain the same result highlighting the strength of our model. Following the DL torque normalization, we can perform the same analysis on the FL torques as well. We obtain the same result that there is no intrinsic increase from the oxidation of Pt, as plotted in Figures 3 (c) and (d), with overall domination of DL-SOT compared to FL-SOT in agreement with previous reports for Co/Pt interfaces. [38]

Another method of verifying this model is by considering the changes that would occur due to interfacial oxidation. In such a case, the oxygen present at the interface would affect the Rashba field at this interface. This would in turn cause the FL torques to change as they are strongly correlated with the Rashba effect. [39,40] Hence, with increasing oxidation, we would expect FL torque to change significantly compared with the DL torque. However, the opposite is observed wherein the DL and FL torques have the same thickness dependence with Pt. This is plotted in Figure 3 (e) and (f) respectively. This strengthens our model of metallic Pt interface and oxygen diffusion within the Co layer.

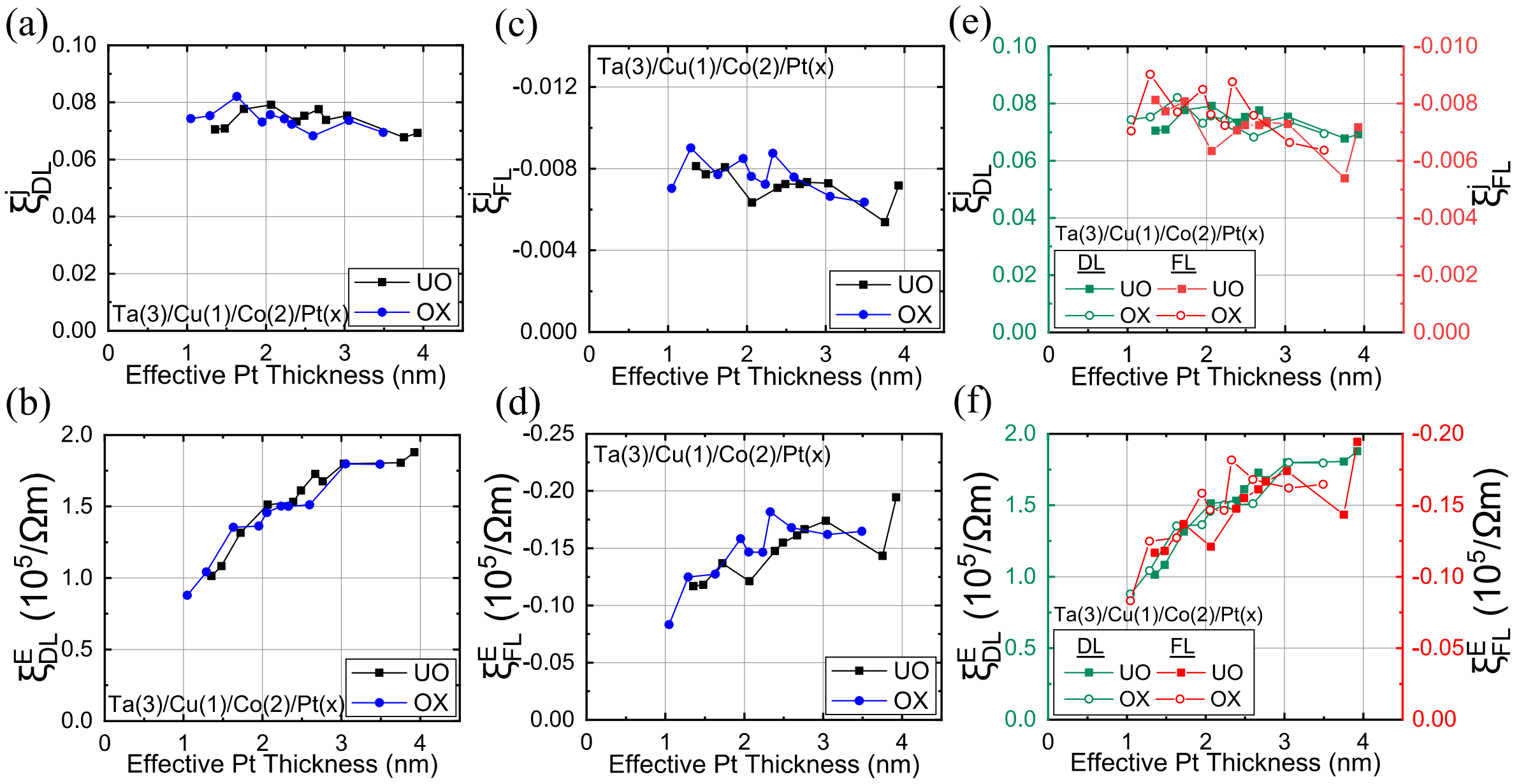

**Figure 3.** DL SOT generation efficiency normalized by (a) the current density and (b) the applied electric field. FL SOT generation efficiency normalized by (c) the current density and (d) the applied electric field. Comparison of the DL SOT efficiency and the FL SOT efficiency normalized by (e) the current density and (f) the applied electric field.

## 5. Conclusion

In conclusion, an apparent increase in torques upon oxidizing platinum is observed, similar to other works. However this apparent increase disappears when considering a new scenario of HM oxidation in Co/Pt where the HM is no longer completely oxidized but instead oxygen is pumped into the FM, leaving the HM metallic at the interface. This oxygen further diffuses into the bulk of the FM layer affecting its magnetic and electronic properties. These changes of the FM poperties are solely responsible for the apparent SOT increase. After appropriate normalizations with resistivity and magnetization we observe no intrinsic increase in torques from the oxidation of the HM. This scenario is validated using ab-initio calculations and $M_S$ and anisotropy measurements. This is likely a general effect: when two oxidized materials are placed in contact with each other, depending on their respective oxide enthalpies of formation, one material could pump oxygen into the other. This needs to be considered in oxidized heterostructures, which are increasingly used in MRAM based applications, in order to avoid erroneous conclusions about key parameters.

**Acknowledgements**

J. Nath, E. Chan, H. Yang, E. Schmoranzerova, M. A. Nsibi, and I. M. Miron acknowledge funding for this work from the European Research Council (ERC) under the European Union's Horizon 2020 research and innovation program (Grant Agreement No. 638653 – Smart Design). M. S. Gabor acknowledges funding from MRI-CNCS/UEFISCDI through grant PN-III-P4-ID-PCE-2020-1853.

# Supporting Information

**Mechanism of Spin-Orbit Torques in Platinum Oxide Systems**

*Jayshankar Nath*, Alexandru Vladimir Trifu, Mihai Gabor, Ali Hallal, Stephane Auffret, Sebastien Labau, Aymen Mahjoub, Edmond Chan, Avinash Kumar Chaurasiya, Amrit Kumar Mondal, Haozhe Yang, Eva Schmoranzerova, Mohamed Ali Nsibi, Isabelle Joumard, Anjan Barman, Bernard Pelissier, Mairbek Chshiev, Gilles Gaudin, and Ioan Mihai Miron**

**Contents**

## S1. Film Preparation

The films were deposited on a 100 mm wafer using a DC magnetron sputtering system from Actemium with a base pressure of $8\times10^{-6}$ Pa and in an Ar atmosphere. The substrate is displaced by 100 mm from the target axis to obtain a wedge of Pt, as shown in **Figure S1**. The deposited samples are immediately protected with a PMMA resist to avoid spurious oxidation of the samples and the film thicknesses are determined using conductance of calibration samples deposited on- and off-target axis, as shown in **Figure S2** (a). Here, the flat edge of the off-axis wafer is taken as the origin of the x-axis, such that the target axis is at -50.75 mm. The on-axis samples give us access to the conductance of the nominal on-axis thickness of platinum, which is known. Each unique conductance value of the off-axis sample corresponds to a unique platinum thickness. Hence, to determine the thickness of the wedge of platinum in a large thickness range, we need to find a fitting function representing the thickness variation versus the distance, that aligns the different G(S) vs thickness curves together as shown in Figure S2 (b). The deposited wafer is diced as illustrated in **Figure S3**, for the experimental studies.

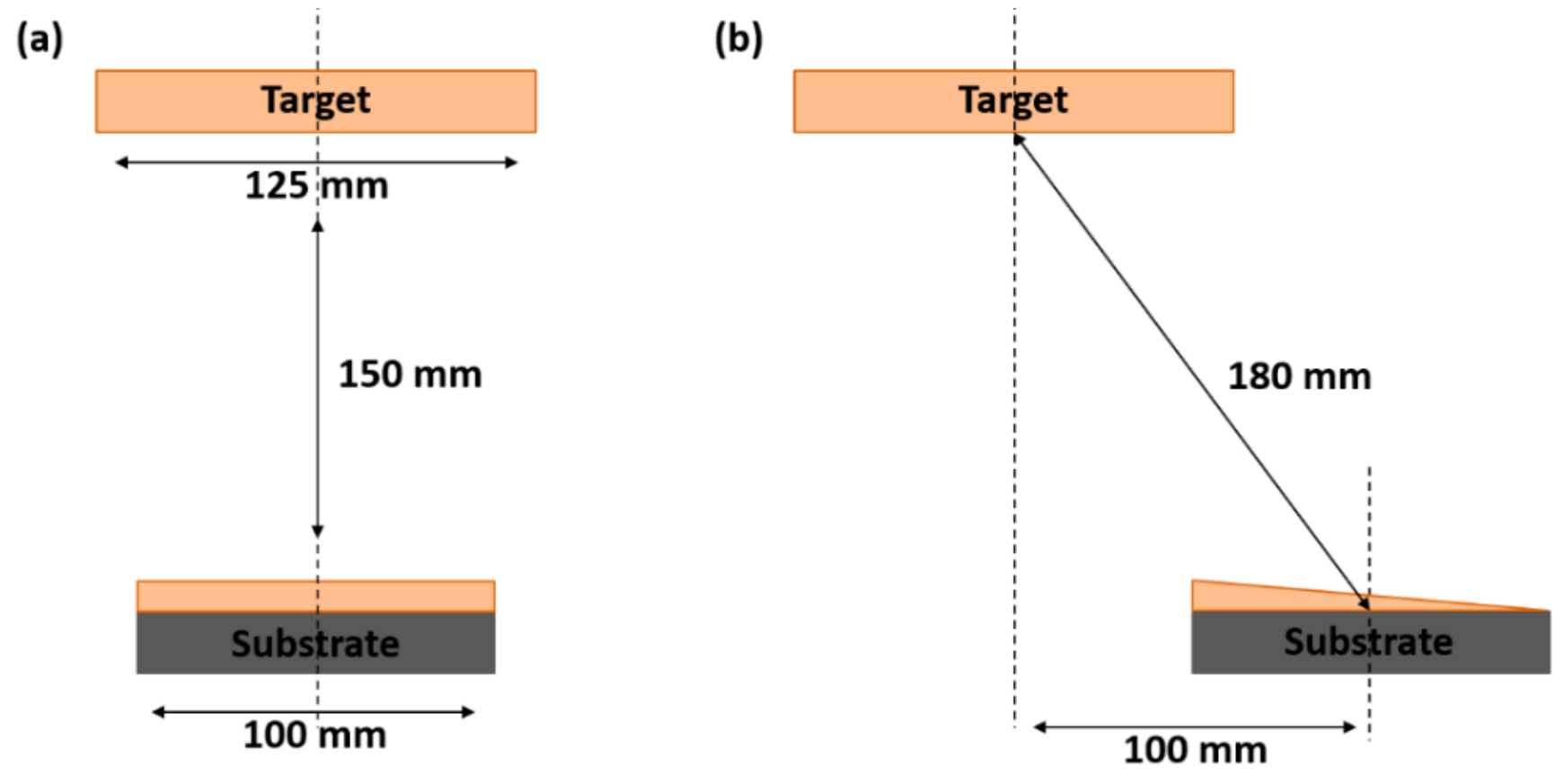

**Figure S1.** Schematics of a) On-axis and b) Off-axis substrate positions to obtain a uniform and a wedge deposition respectively.

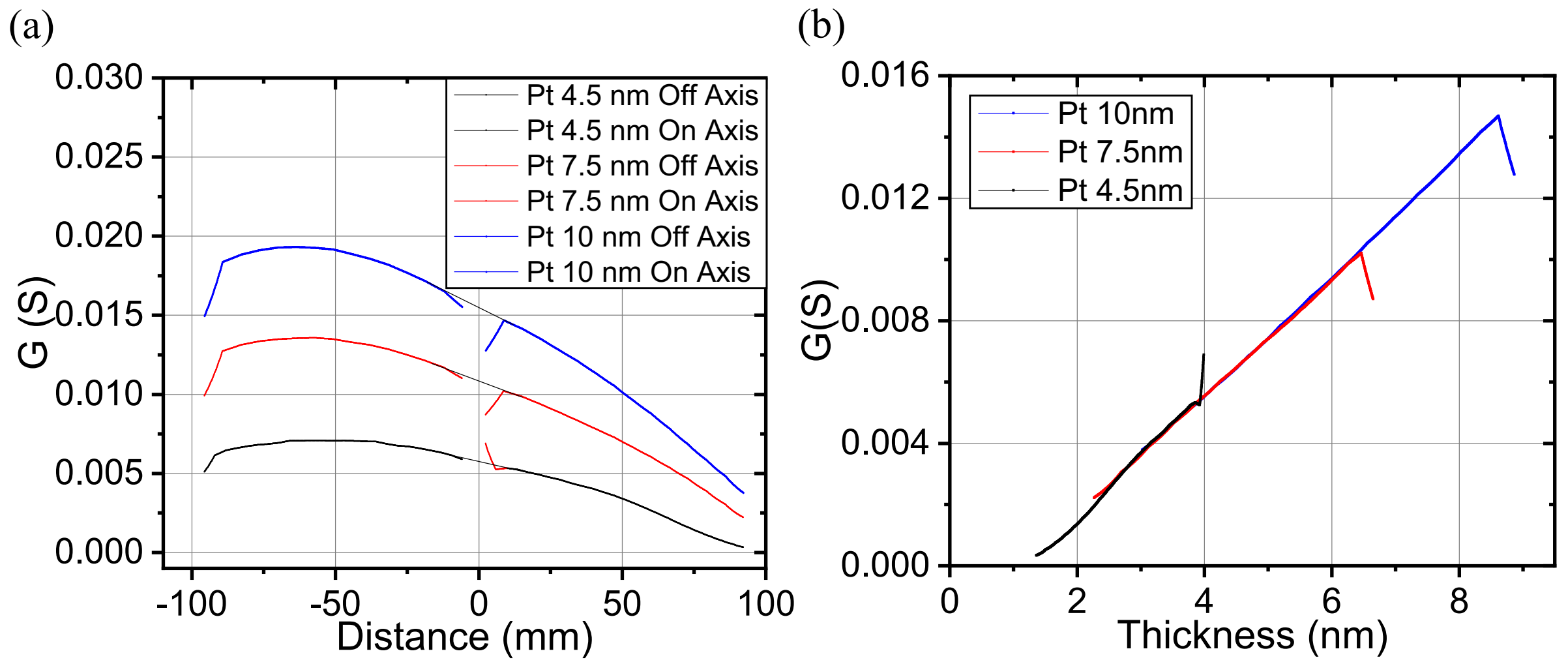

**Figure S2.** a) Conductance of Pt calibration samples grown on- and off-axis with respect to the target. b) The dependence of conductance on the thickness of platinum determined via optimization technique.

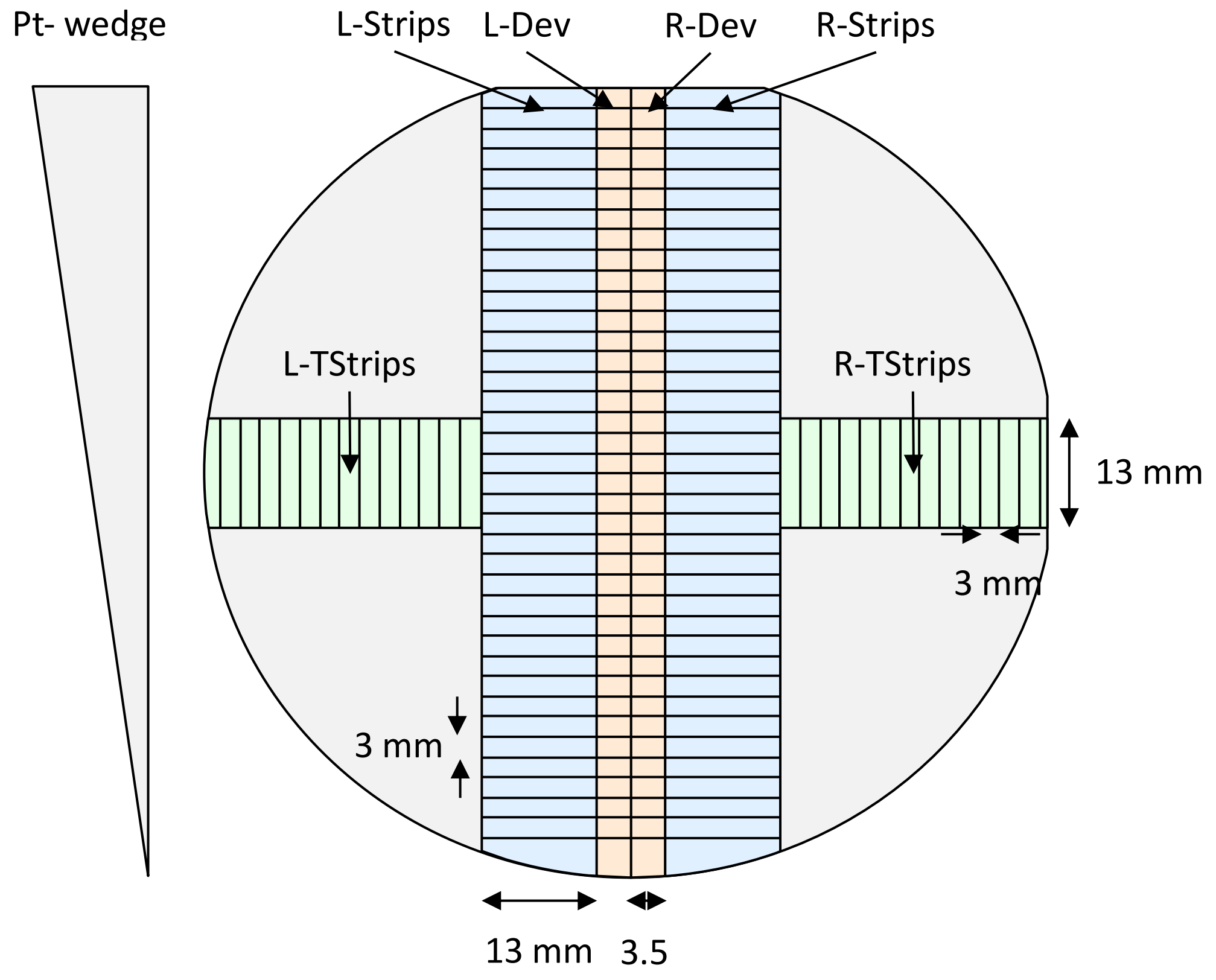

**Figure S3.** Dicing schematic of the wafer, which is laterally symmetric. The platinum wedge direction is shown on the left. Strips in blue, “L-Strips” and “R-Strips,” are diced along the wedge for use in VSM and FMR measurements. The orange strips, “L-Dev” and “R-Dev,” are utilized in torque measurements after device fabrication. The green strips, “L-TStrips” and “R-TStrips,” are diced transverse to the wedge to determine the lateral variation during deposition.

**S2. Device Fabrication**

The samples were patterned into 5 μm wide Hall crosses using UV lithography and Ar ion milling [S1] as shown in the process flow of **Figure S4**. The processed devices were re-coated with PMMA resist to avoid spurious oxidation.

(a)

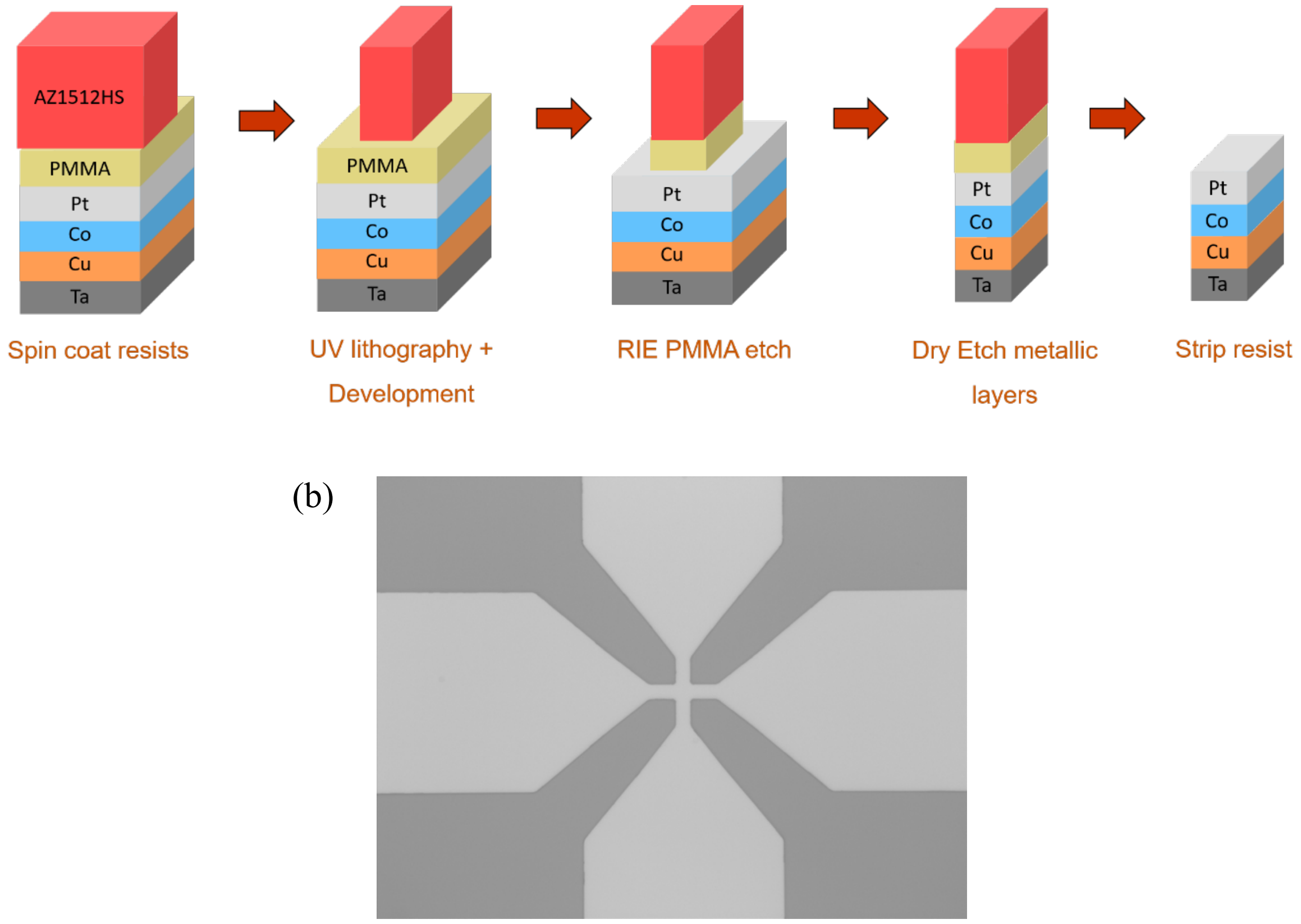

(b)

**Figure S4.** a) Dual resist microlithography process flow. b) Optical micrograph of the Hall cross device with a width of 5μm.

### S3. Platinum Oxidation

The sample oxidation was performed in an Inductively Coupled Plasma – Reactive Ion Etching (ICP-RIE) chamber with an argon-oxygen (20sccm-5sccm) atmosphere. The platter is maintained at 0V to avoid kinetic ion-etching of the sample during oxidation.

### S4. AR-XPS Characterization (pAR-XPS)

The AR-XPS measurements were performed on a Thermo-Fisher Scientific Theta 300 pARXPS system with a monochromatic aluminum anode source at 1486.6 eV and ultra-high vacuum conditions of $3\times10^{-7}$ Pa. The two-dimensional detector integrated in this tool allows performing a parallel acquisition of the photoemission signals at eight emission angles from 20° to 80° with respect to the sample normal, without any tilt of the sample (parallel Angle Resolved XPS).

## S5. Conductance Measurements and Effective Platinum Thickness

To determine the current flow in Pt, it is necessary to obtain the conductance of the Pt layer and the layers underneath. However, the resistivity of thin films depends on the film thickness and is not constant, as in bulk materials. Having a wedge of platinum allows us to determine this thickness dependence. The conductance of UO and OX are determined using four-probe measurements and are plotted in **Figure S5** (a). The conductance increases with increasing Pt thickness and decreases with oxidation. These curves however include an inherent curvature in the thin films arising from deposition, visible at higher Pt thickness. This curvature can be determined from the conductance measurements of samples transverse to the wedge of Pt and is plotted in Figure S5 (b). A renormalized fit of this curve is used to correct the conductance plot along the wedge, as plotted in **Figure S6**. It shows that the conductance has a linear dependence at a larger Pt thickness due to its constant conductivity, while it deviates from this behavior at lower thicknesses.

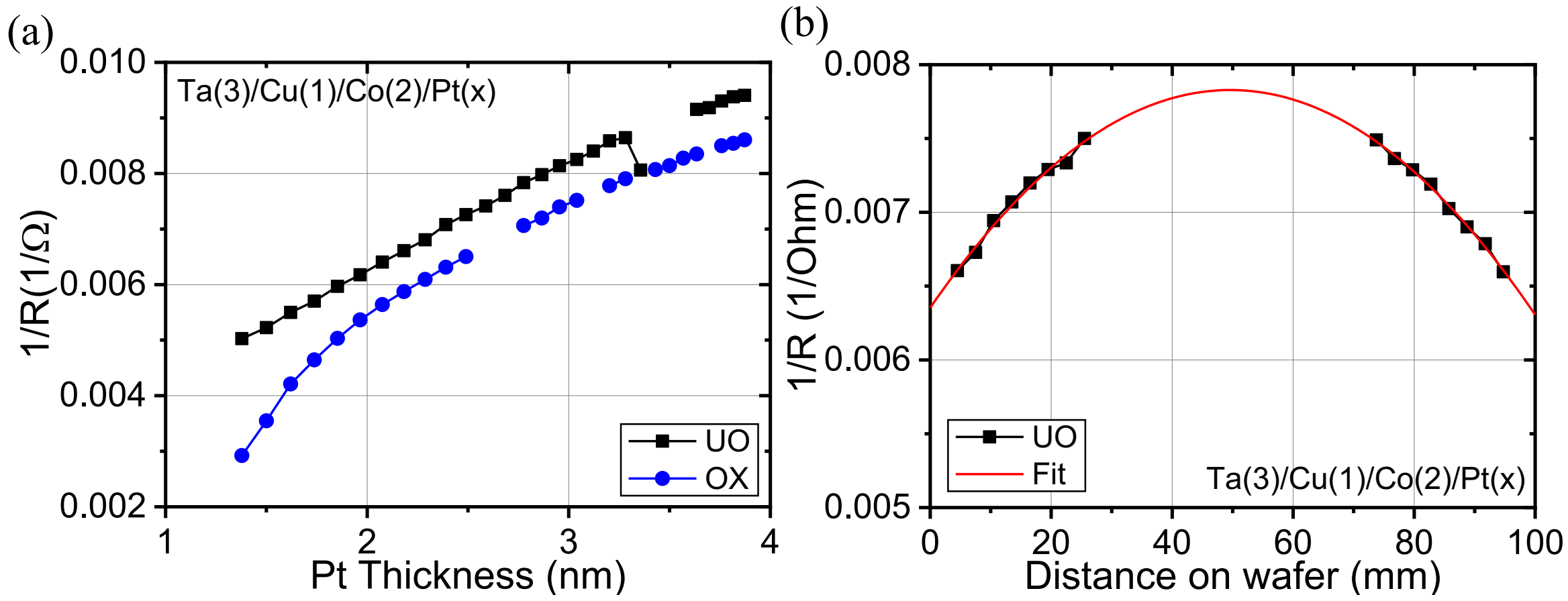

**Figure S5**. (a) Variation of the conductance of the samples with respect to the Pt thickness. (b) The conductance of the UO strips, perpendicular to the Pt wedge, plotted with respect to the transverse position on the wafer.

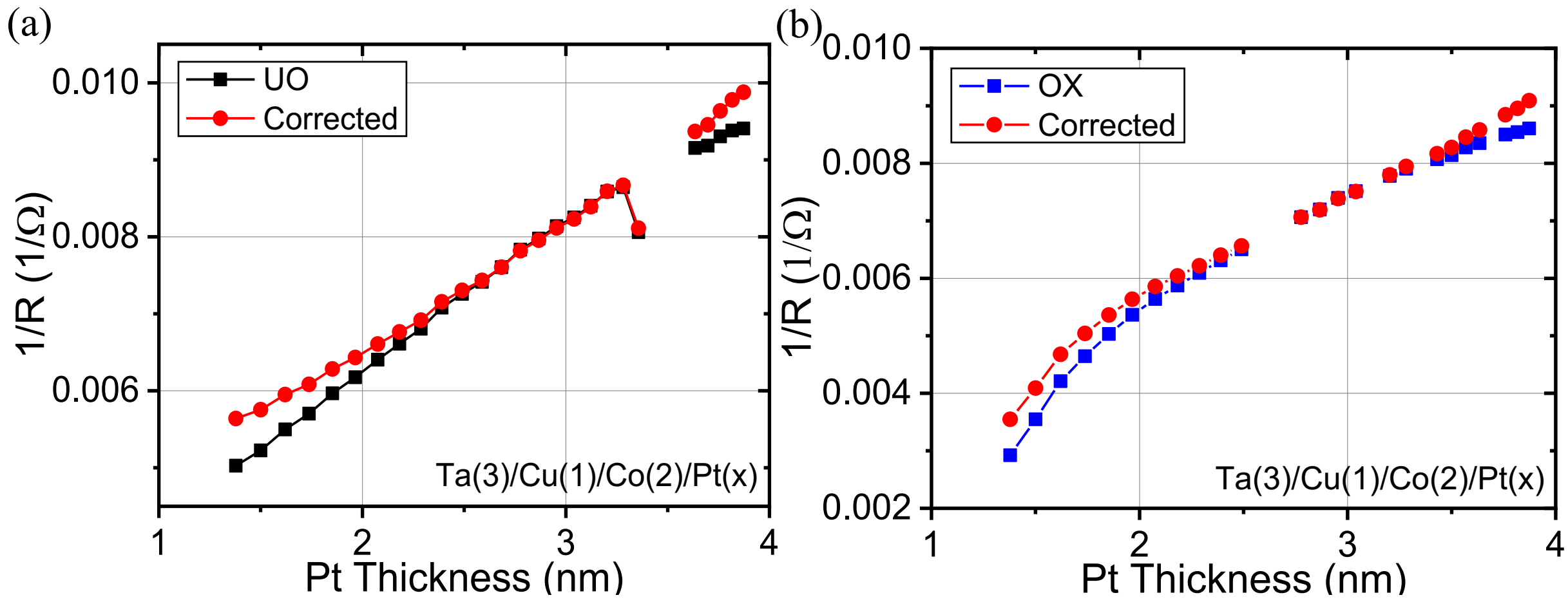

**Figure S6**. The conductance of (a) UO and (b) OX samples corrected for deposition induced curvature.

Following the corrections, we can use the Fuchs-Sondheimer (FS) model [S2, S3] to model the conductance of Pt. As the surface scattering effects should be more prominent, we use a general FS model where the conductance of the film is given by

$$\frac{1}{R} = \frac{1}{R_0} + \frac{w*t}{\rho_0\left(1+\frac{3\lambda}{8t}\right)l} \times (1-p) \tag{S1}$$

Here $R$ is the total resistance of all the layers, $R_0$ the resistance of the uniform layers of the film beneath the Pt layer, $w$ the width of the strip, $t$ the thickness of the platinum layer, $\rho_0$ the bulk resistivity of the platinum layer, $\lambda$ the mean free path of the electrons in the film, $l$ the distance between the voltage probes, and p the specularity parameter. As the surface roughness is larger than the de Broglie wavelength of free electrons, we assume a completely diffuse scattering of electrons from the surface of the metal, leading to $p = 0$. This model is then used to fit the conductance plot of UO, plotted in **Figure S7**. From the fitting, we extracted the values of $R_0$ = 216.46 Ω, $\rho_0$ = 19.42 μΩcm and $\lambda$ = 9.34 nm. Even though the extracted mean free path is larger than the film thickness, the model fits our data quite well and the extracted values are consistent with the literature.[S4] Furthermore, we are more interested in the value of $R_0$, which determines the current flow in the Pt layer and the layers underneath, rather than specific Pt material parameters. The resistance of the Pt layer is given by

$$R_{Pt}(t) = \frac{\rho_0\left(1+\frac{3\lambda}{8t}\right)l}{w*t} \tag{S2}$$

And the resistivity is given by

$$\rho_{Pt}(t) = \rho_0\left(1 + \frac{3\lambda}{8t}\right) \tag{S3}$$

These, along with the conductance of the Pt layer are plotted in Figure S7.

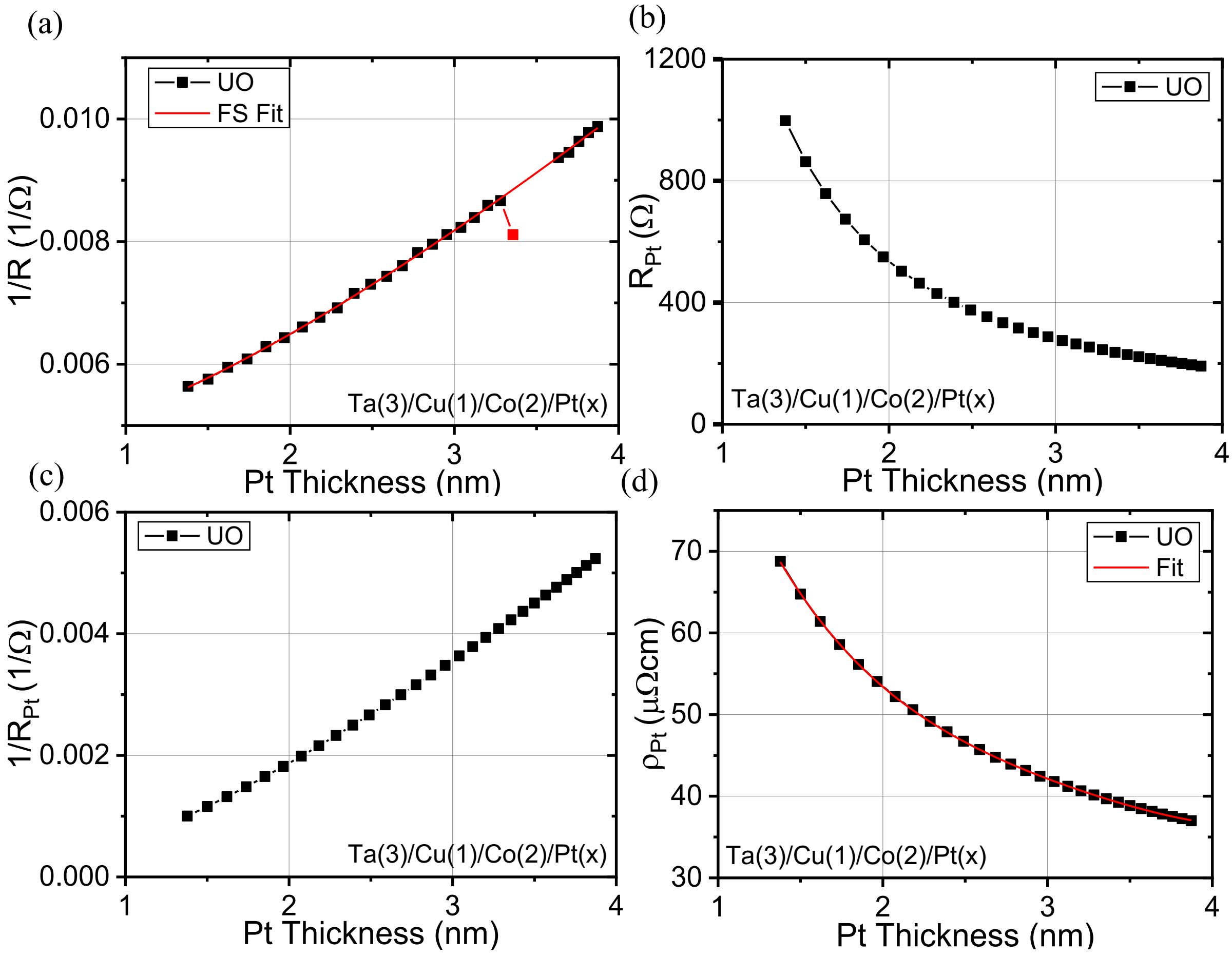

**Figure S7**. (a) Variation of the total conductance of the UO samples with respect to the Pt thickness. The red curve is the FS fit to the data. Dependence of (b) the resistance, (c) the conductance, and (d) the resistivity of the platinum layer on its thickness. The polynomial fit to the data is plotted by the red curve in the last figure.

At larger thicknesses of Pt, there is a constant offset between the OX and UO samples visible in the conductance plot of Figure S5. This constant offset can be considered as the reduction of the effective thickness of the electrically conducting platinum since the plasma oxidation of platinum oxidizes a uniform layer of platinum into platinum oxide.[S5] This is visualized in Figure 2 (c) of the main text. The thickness of this oxide layer hence corresponds to the lateral offset of the conductance plot of OX, as plotted in **Figure S8** (a). Correcting for this offset, the conductance of UO and OX with respect to the effective Pt thickness is plotted in Figure S8 (b).

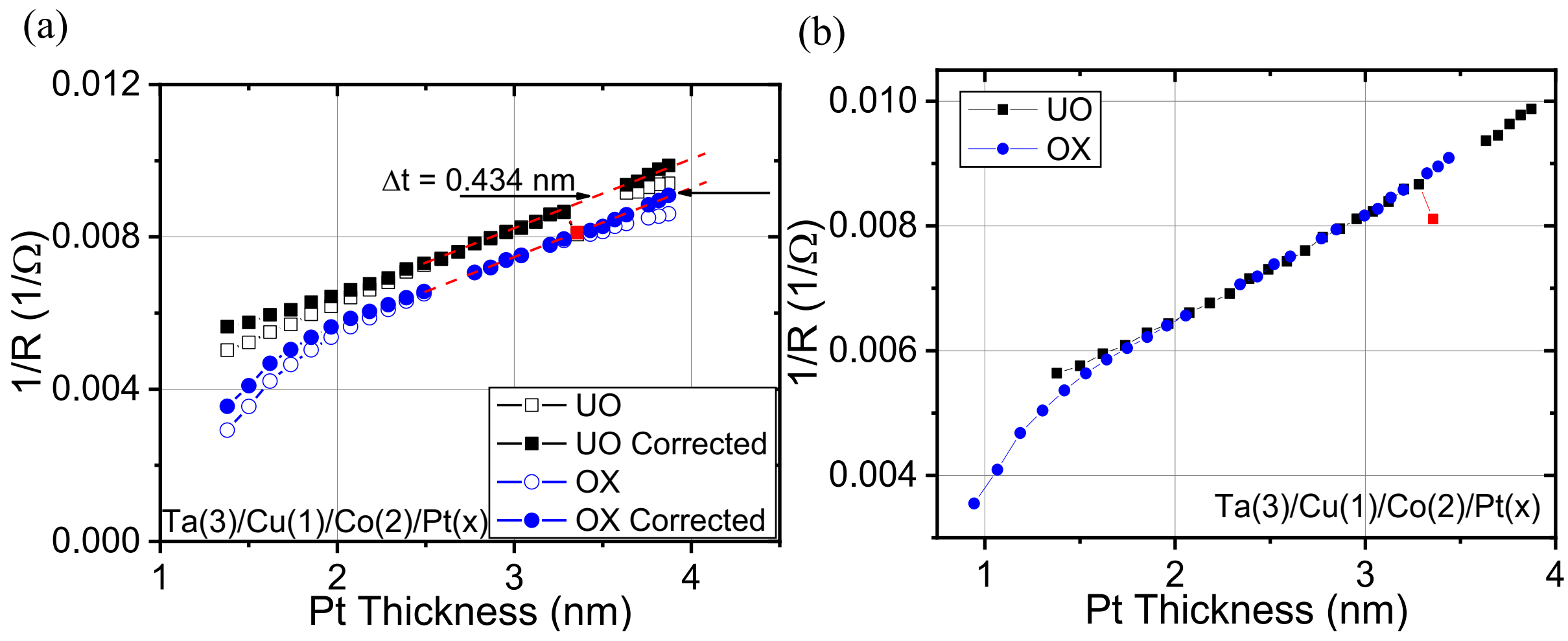

**Figure S8**. (a) Determination of the platinum oxide thickness from the lateral offset of conductance plots. Here, the corrected curves refer to the deposition induced layer curvature correction. (b) The conductance plots of UO and OX with respect to the effective platinum thickness.

**S5. Magnetic Characterization and Co Thickness Correction**

The VSM measurements were performed on a Microsense VSM at room temperature. The total magnetization of the sample at different Pt thicknesses is plotted in **Figure S9** (a). There is a symmetric curvature present in the magnetization plot of UO samples. This arises from the deposition process as seen in the previous section on sample resistances and needs to be corrected. This can be achieved by considering the fact that at the center of the wafer, the Co is 2 nm thick, fixed by the calibration of the deposition system, and it tapers towards the edge of the 4-inch wafer. Ideally, as the saturation magnetization of the Co layer is constant, the curvature of the magnetization plot arises primarily from this thickness variation of the Co and can be extracted as shown in Figure S9 (b). Correcting for this thickness variation, we can obtain the saturation magnetization of the samples as plotted in the main text. As for the OX sample, the drop in magnetization at lower thicknesses of Pt could arise from the oxygen diffusion into Co resulting in a drop in the $M_S$ value without affecting the Co thickness. It could also result from oxygen forming an interfacial layer between Pt/Co. In this case, the $M_S$ would have been constant while resulting in a drop in Co thickness. In the manuscript, the former is shown to be the case by comparing the interfacial anisotropy of the samples. The Co thickness variation needs to be accounted for, in the determination of interfacial anisotropy. Hence, the anisotropy values calculated in Equation (1) of the main text are multiplied by a thickness correction factor of (Co thickness/nominal Co thickness (2 nm)) for each sample and is plotted in Figure 2 (e) of the main text.

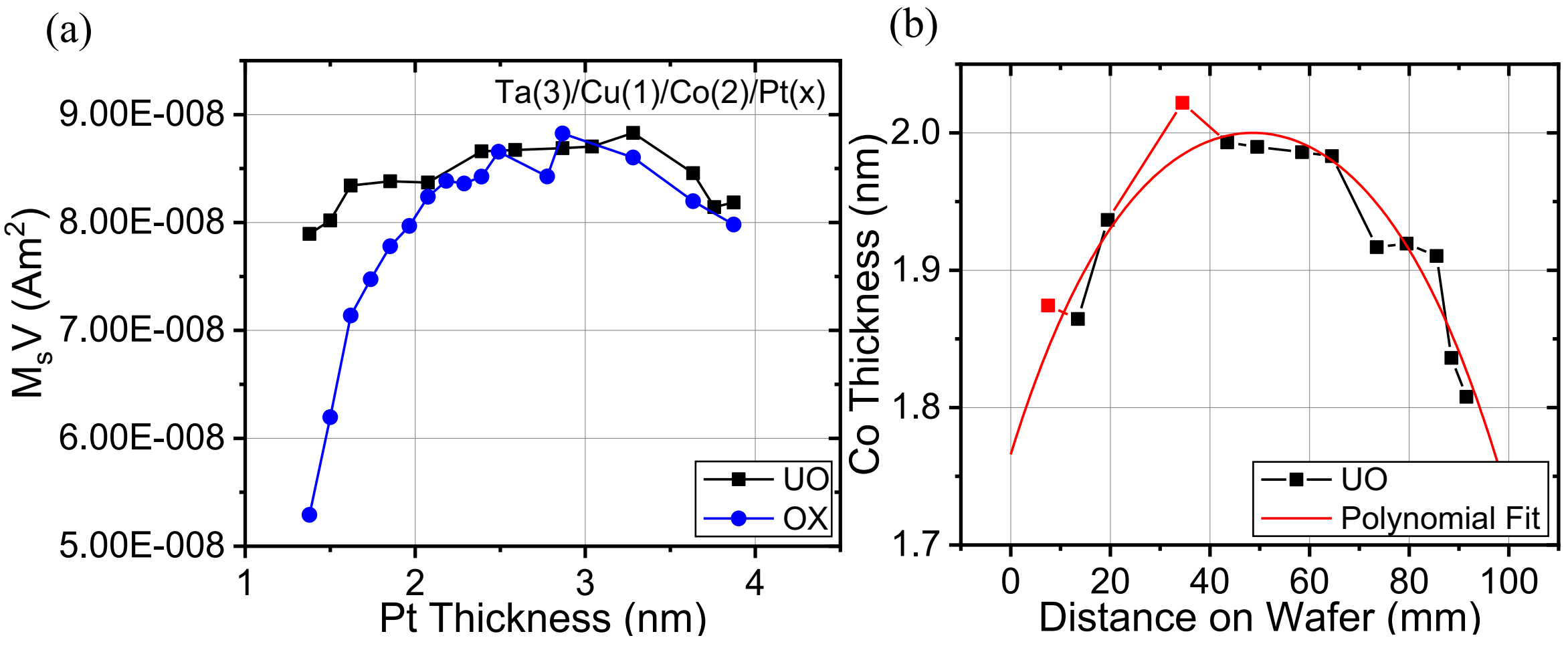

**Figure S9**. (a) The total magnetization of the samples along the Pt wedge. (b) Co thickness along the wafer. On-axis wafer deposition results in a drop in thickness towards the edges of the sample.

### S6. SOT Measurements

The effect of current-induced SOTs needs to be compared with that of an effective field to be able to quantify it. This can be achieved by measuring the effect of each on the magnetization of the sample, as shown in **Figure S10**. In practice, a Hall measurement setup is used with a small AC current of 10 Hz applied along one of the arms of the Hall cross. The transverse AC voltage is measured along the other, as shown in **Figure S11**. A linear response would result in no oscillation of the magnetization. A non-linear response is indicative of magnetic oscillations, and hence of SOTs, and would give rise to higher harmonics.

(a) 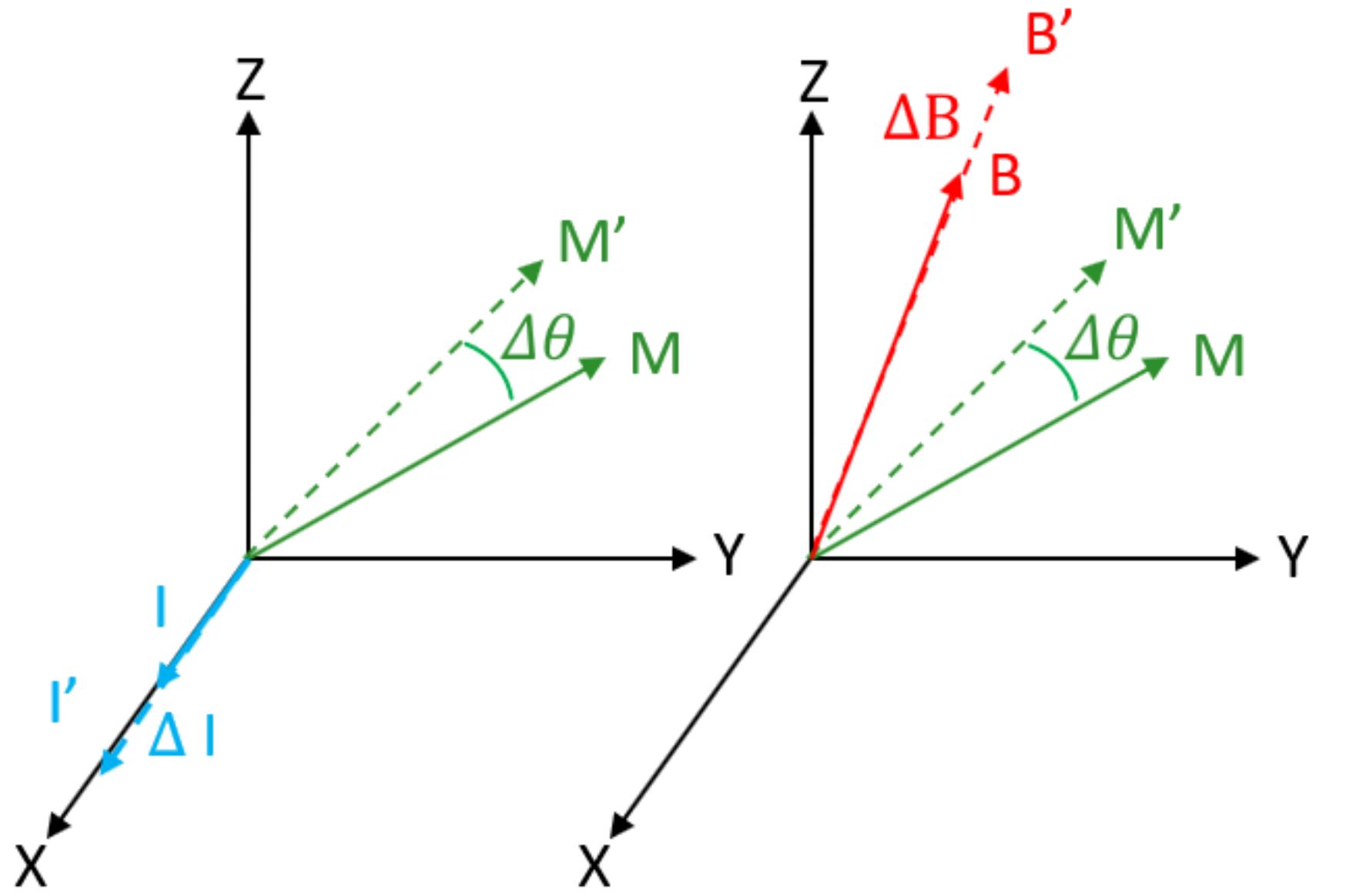 (b)

**Figure S10**. Working principle of the 2nd harmonic torque measurements. The variation of magnetization (a) caused by the applied AC current is compared (b) to that caused by an externally applied field.

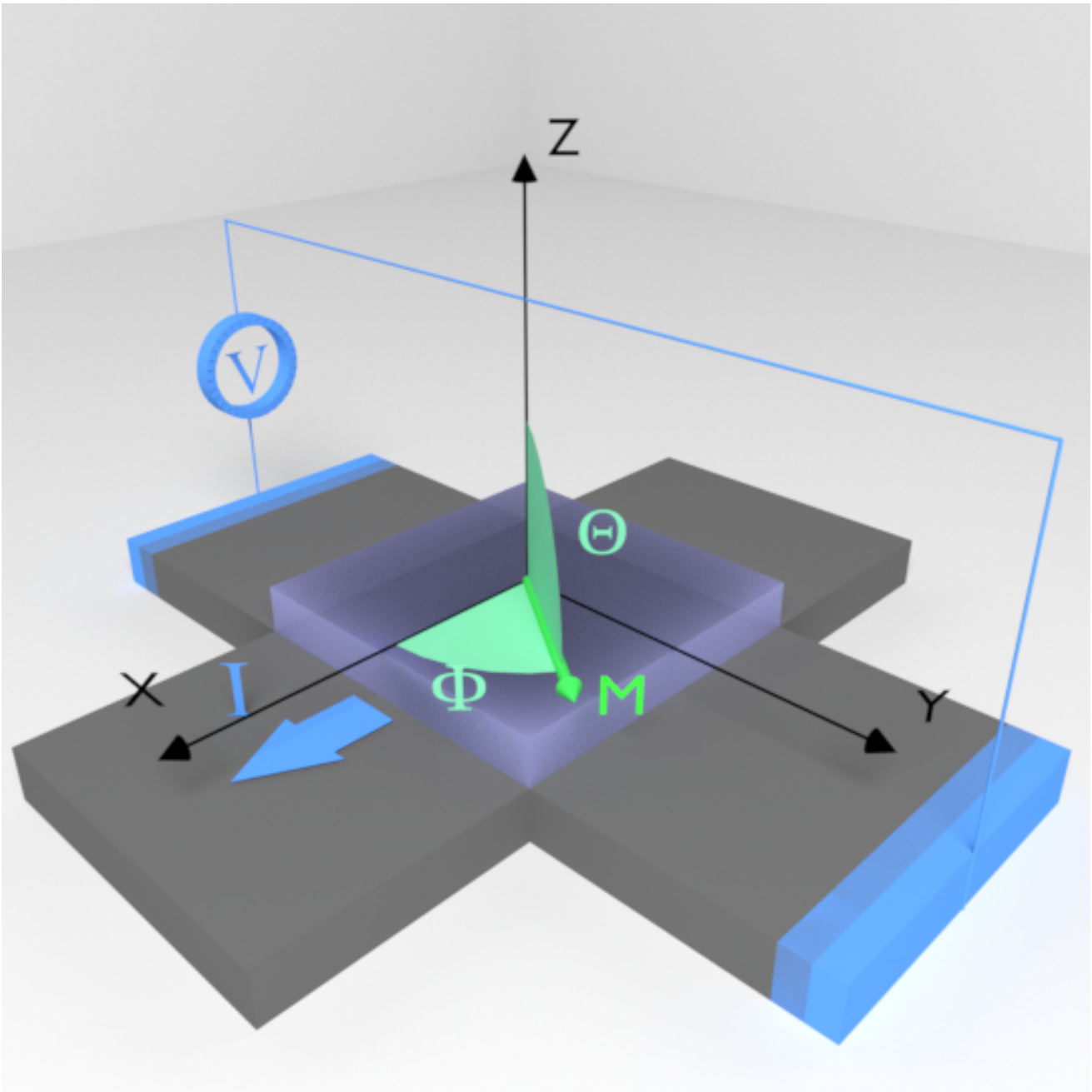

**Figure S11**. Hall measurement setup for 2nd harmonic torque measurements. A small AC current is applied along one set of the arms of the Hall cross while the voltage is measured transverse to the current. The magnetization of the sample can be rotated using an external field.

As this work deals with in-plane magnetized materials, the external field is rotated in the plane of the sample during the measurement. The first harmonic signal provides the measure of the Planar Hall Effect (PHE), indicative of the position of magnetization, as plotted in **Figure S12** (a). The second harmonic signal plotted in Figure S12 (b) denotes the oscillations of the magnetization arising from the non-linearities, such as DLTs, FLTs, and thermal signals. These signals are damped as the field strength is increased, indicating the stiffening of the magnetization.

The second harmonic signals can be separated into their constituent components based on the angular symmetry given by [S1, S6-S8]

$$R_{DL}^{2f} + R_{\Delta T}^{2f} \sim \cos\varphi \quad \text{(S4)}$$

$$R_{FL}^{2f} \sim 2\cos^3\varphi - \cos\varphi \quad \text{(S5)}$$

Here, $R_{DL}^{2f}$ and $R_{FL}^{2f}$ denote the components of DLT and FLT respectively. $R_{\Delta T}^{2f}$ is the thermal component which arises due to the thermal gradient induced by the current injection into the sample. $\varphi$ is the in-plane angle of the magnetization. These are plotted in **Figure S13** (a) and (b) respectively.

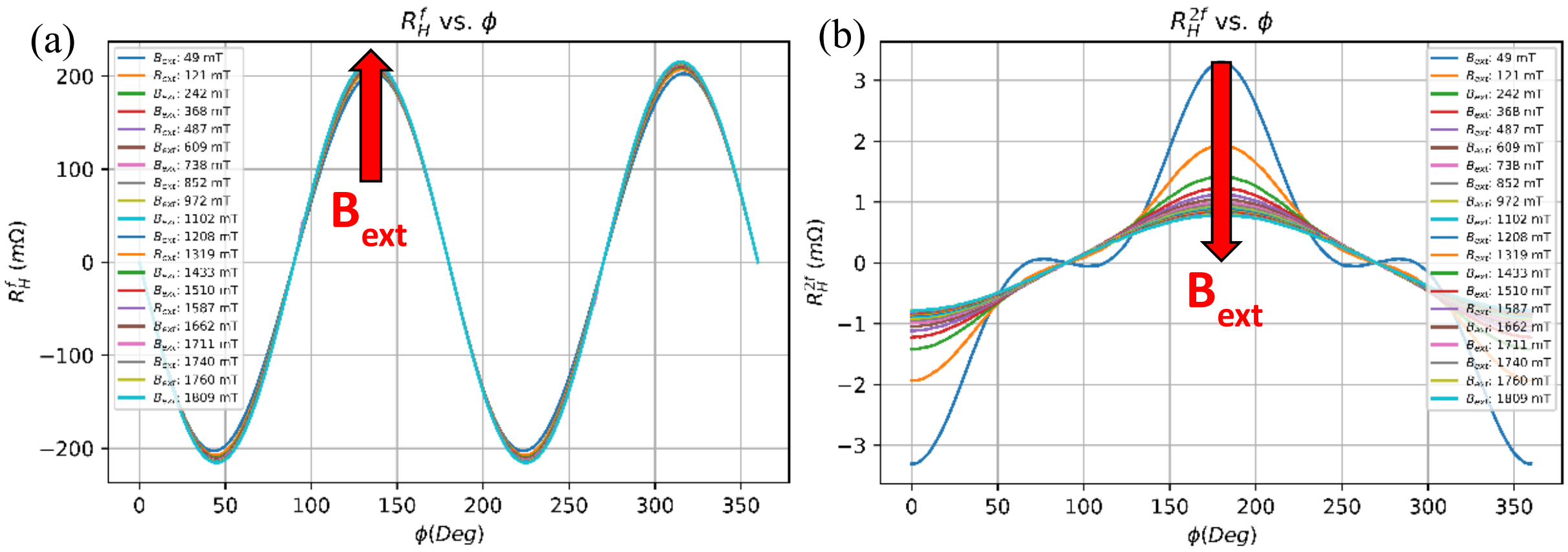

**Figure S12.** In-plane angular dependence of (a) the first and (b) the second-harmonic Hall resistances, measured from an in-plane angular scan of Ta(3)/Cu(1)/Co(2)/Pt(3).

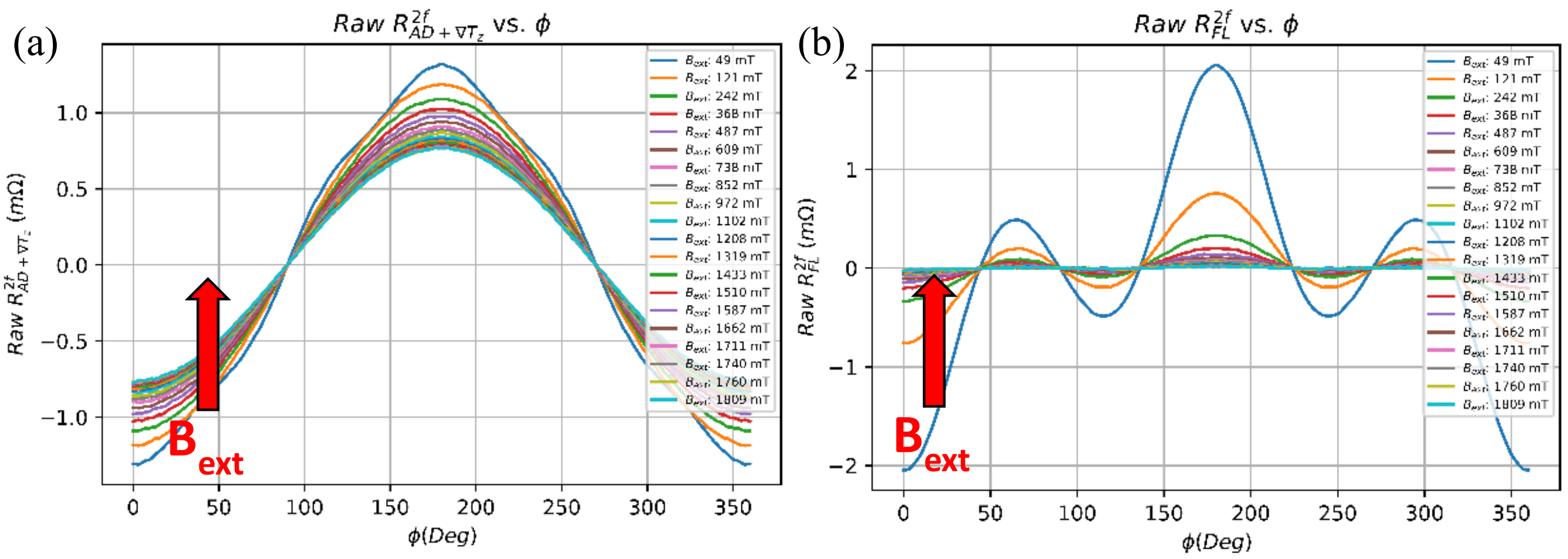

**Figure S13.** In-plane angular dependencies of (a) the DL and the thermal, and (b) the FL components.

The DL and FL fields vary inversely with the total field acting on the magnetization promoting or retarding its motion along their respective directions. They can hence be extracted and quantified from the slopes of these curves, plotted in **Figure S14**, and given by

$$B_{DL} = -\frac{slope}{R_{AHE}} \tag{S6}$$

$$R^{2f}_{\nabla T} = y - intercept \tag{S7}$$

$$B_{FL} - B_{Oe} = \frac{slope}{2R_{PHE}} \tag{S8}$$

Here $B_{DL}$ and $B_{FL}$ are the DL and FL fields. $B_{Oe}$ is the Oersted field arising from the non-magnetic layers of the sample and $R_{AHE}$ is the Anomalous Hall contribution to the Resistance. (S6) and (S7) are calculated based on Figure S14 (a) and (S8) based on Figure S14 (b).

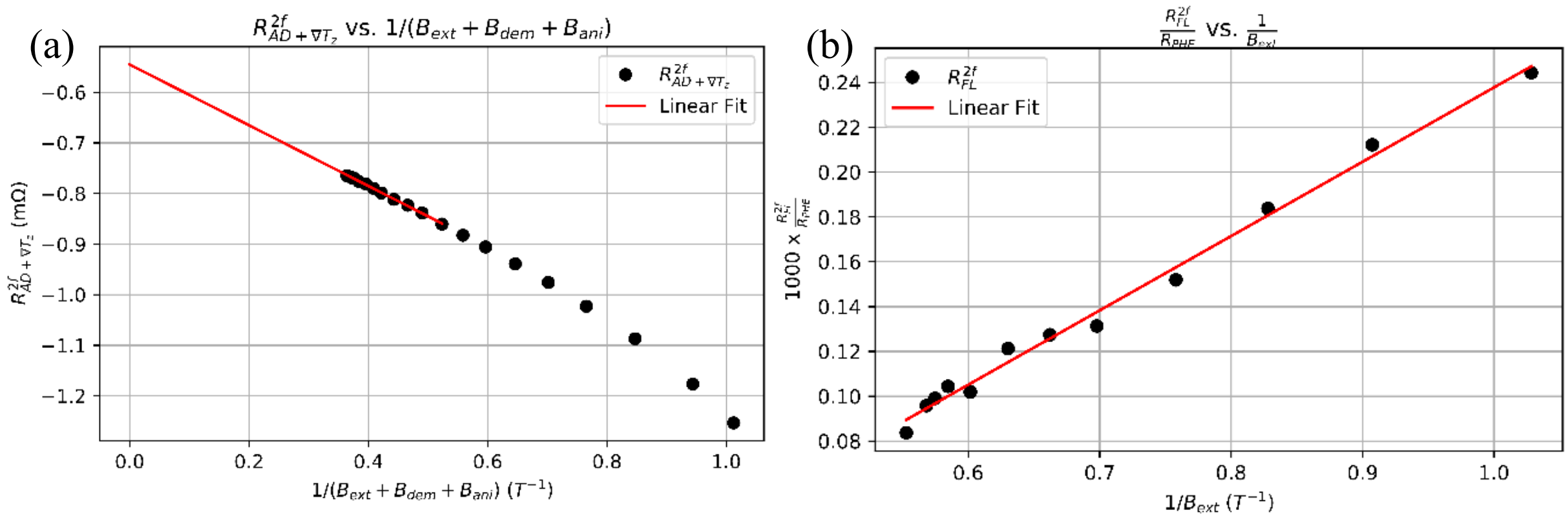

**Figure S14.** Dependence of the (a) the DL and the thermal, and (b) the FL component of the second harmonic resistance on the inverse field acting on the magnetization.

**S7. FMR Measurements**

The FMR measurements were performed using a microwave TE 011 cavity operating in the X band (9.79 GHz) with a power of 1 mW. An external field H modulated by an AC field is swept across the sample and the derivative of the microwave absorption spectra provides the FMR spectra. The dependence of the resonant field and the linewidth on the external field angle are plotted in **Figure S15**. The anisotropy fields of the samples are extracted from the resonance field plot. The larger resonant field at lower field angles denotes an in-plane anisotropy in our samples.

The external field dependence of the linewidth provides access to the damping parameter α, plotted in **Figure S16** (a). All the samples have a higher damping constant when compared with the reference sample, REF. This can be attributed to the higher SOC of the platinum layer. The damping constant is even higher in the case of the oxidized samples. The g-factor, given by $\frac{\mu_L}{\mu_S} = \frac{g-2}{2}$, relates the spin ($\mu_S$) and orbital ($\mu_L$) components of the net magnetic moment and is plotted in Figure S16 (b). At interfaces where the crystal symmetry is broken, the orbital component has a larger contribution. This is especially so, in the case of a HM with a large SOC. However, the orbital angular momentum can be quenched as a result of the enhanced crystal field due to oxygen. This decreases the g-factor, as visible in the plot. Subsequently, the interfacial spin mixing conductance is expressed in terms of the enhancement of the damping constant Δα as follows [S9] and is plotted in Figure 1(e) of the main text.

$$g_{eff}^{\uparrow\downarrow} = \frac{4\pi M_s d}{g\mu_B}\Delta\alpha \qquad \text{(S9)}$$

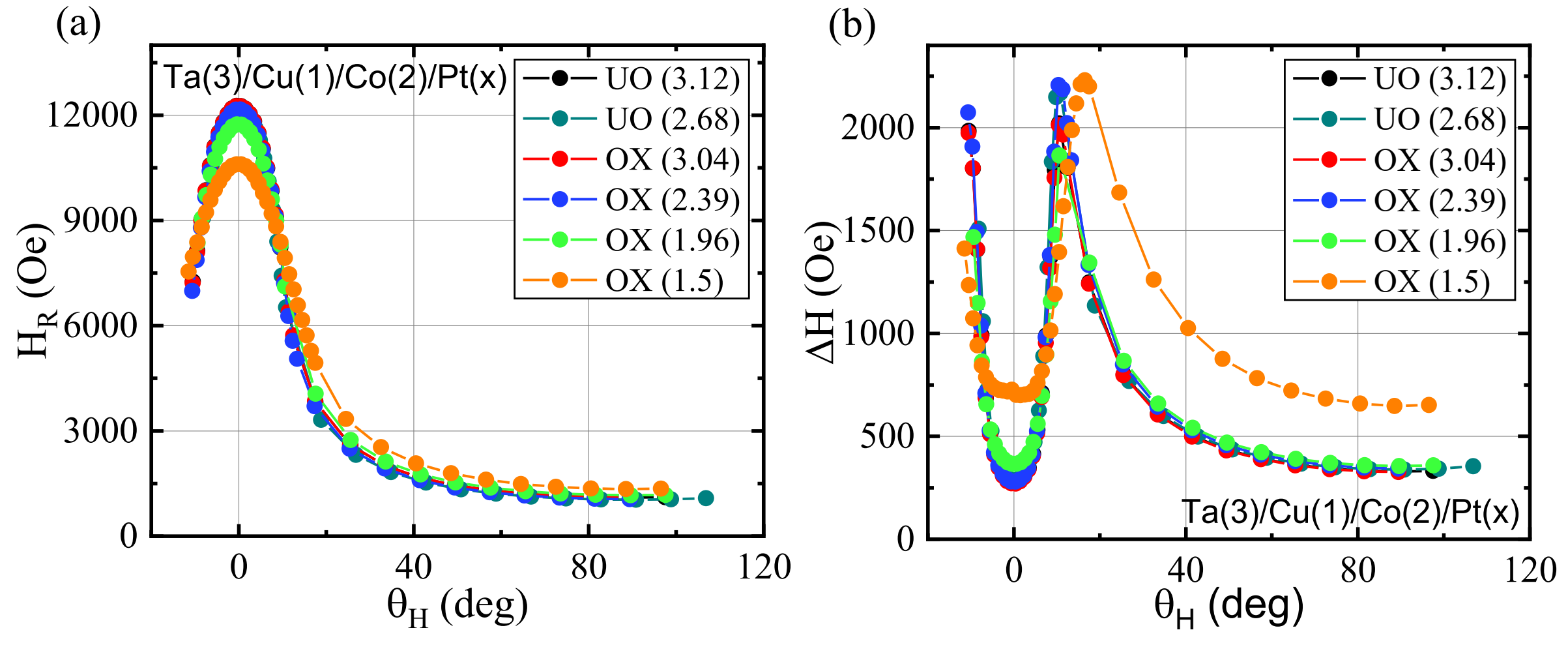

**Figure S15.** The dependence of (a) the resonant field $H_R$ and (b) the linewidth ΔH on the external field angle $\theta_H$. The numbers in the parenthesis indicate the thickness of the top Pt layer in nm.

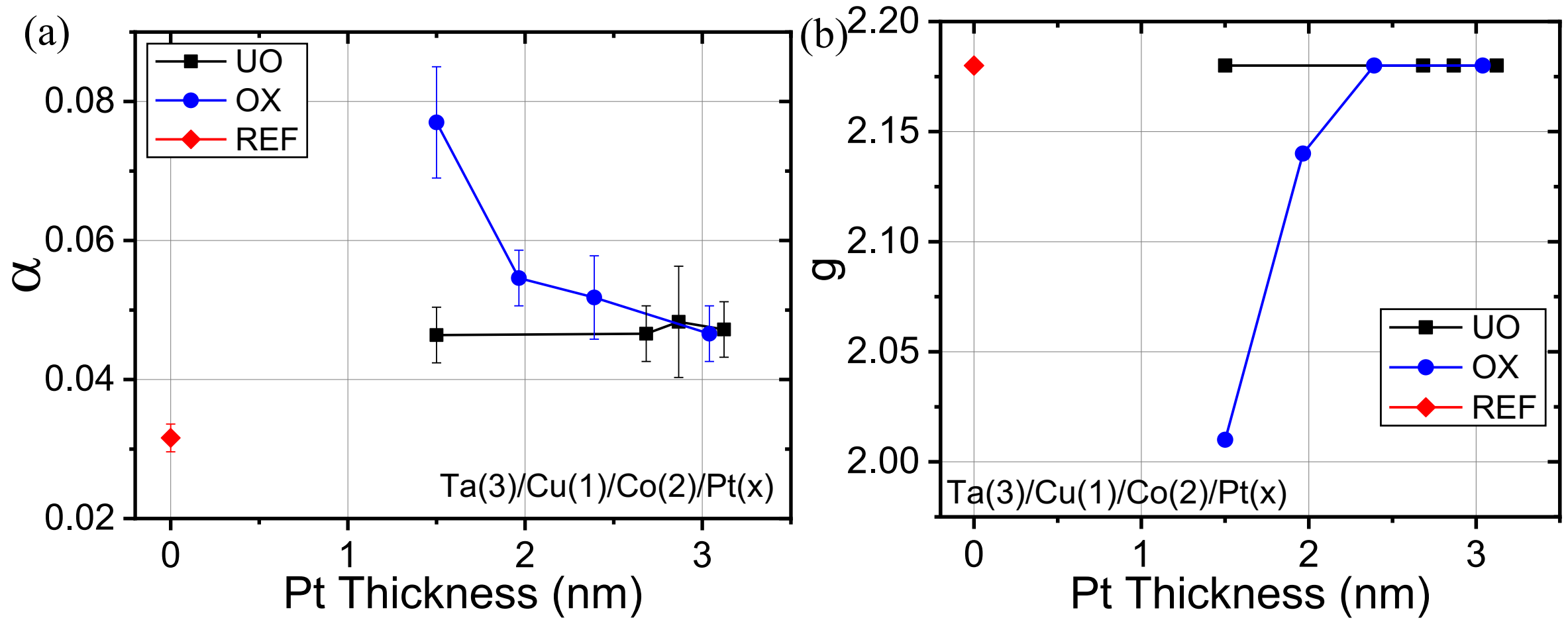

**Figure S16.** Dependence of (a) the damping constant α and (b) the g-factor on the top platinum thickness. REF refers to a reference sample without a Pt layer, Ta (3 nm)/Cu(1 nm)/Co(2 nm)/Al(2 nm). Samples UO (1.5, 2.9) were deposited separately.

**S8. Ab-initio DFT calculations**

The ab-initio calculations to determine the energetics of the system as oxygen atoms are placed into the lattice, shown in Figure 2 (d) of the main text, were implemented in the Vienna Ab-initio Simulation Package (VASP) [S10-S12] using the Generalized Gradient Approximation. [S13] The system consists of Pt(3ML)/Co(3ML)/Pt(5ML) tri-layer structure as plotted. An 11 x 11 x 1 k-point mesh was used to sample the first Brillouin zone and a kinetic energy cut-off of 520eV was used for the plane-wave basis set. A plane of oxygen atoms was inserted into the lattice along the X-Y plane and allowed to relax to determine the energy of the system. This plane of atoms was then displaced along the Z direction to determine the system energy as the oxygen is placed at different locations in the lattice. The in-plane lattice constant was fixed at 2.115Å, similar to that of Pt, and the structure was relaxed until the forces were smaller than 1 me/Å.